Research Article

# Bionic Optical Physical Unclonable Functions for Authentication and Encryption

Yongbiao Wan[†, ‡, #], Pidong Wang[†, ‡, #], Feng Huang[†, ‡], Jun Yuan[†, ‡], Dong Li[†, ‡], Kun Chen[†, ‡], Jianbin Kang[†, ‡], Qian Li[†, ‡], Taiping Zhang[†, ‡], Song Sun[†, ‡], Zhiguang Qiu[§, *], Yao Yao[†, ‡, *]

[†] Microsystem & Terahertz Research Center, China Academy of Engineering Physics, Chengdu 610200, China

[‡] Institute of Electronic Engineering, China Academy of Engineering Physics, Mianyang 621999, China

[§] School of Electronics and Information Technology, State Key Lab of Opto-Electronic Materials & Technologies, Guangdong Province Key Lab of Display Materials and Technologies, Sun Yat-sen University, Guangzhou 510275, China.

* To whom correspondence should be addressed. Email: zhiguangqiu88@gmail.com; yaoyao_mtrc@caep.cn

[#] These authors contributed equally to this work.

## ABSTRACT

Information security is of great importance for modern society with all things connected. Physical unclonable function (PUF) as a promising hardware primitive has been intensively studied for information security. However, the widely investigated silicon PUF with low entropy is vulnerable to various attacks. Herein, we introduce a concept of bionic PUFs inspired from unique biological architectures, and fabricate four types of bionic PUFs by molding the surface micro-nano structures of natural plant tissues with a simple, low-cost, green and environmentally friendly manufacturing process. The laser speckle responses of all bionic PUFs are statistically demonstrated to be random, unique, unpredictable and robust enough for cryptographic applications, indicating the broad applicability of bionic PUFs. On this ground, the feasibility of implementing bionic PUFs as cryptographic primitives in entity authentication and encrypted communication is experimentally validated, which shows its promising potential in the application of future information security.

**Graphical abstract**

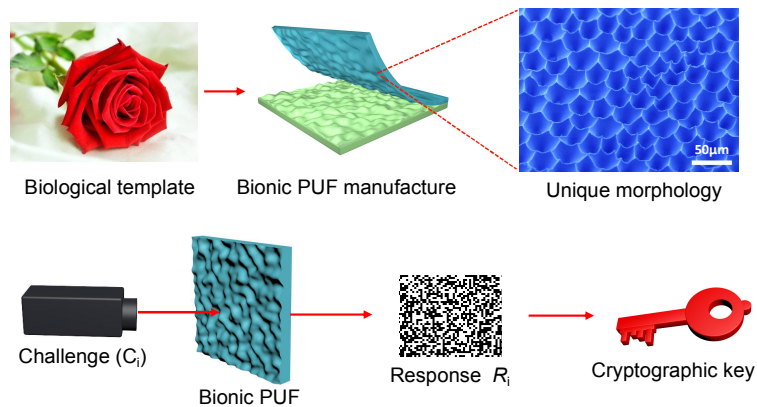

# 1. INTRODUCTION

Remarkable advances have been witnessed in the field of information and communication technologies such as the Internet of Things, big data, cloud computing and artificial intelligence in recent years.[1-6] Such technologies have spawned an extremely complex digital network, connecting physical objects of modern society. The exponential growth of connected physical entities with oceans of data being produced every day poses severe challenges to information security. Even a few caught off guard cyber-attacks could lead to a huge number of data leakages from governments and private organizations worldwide, which exposes the vulnerability of present state-of-the-art information security systems.[7] Typically, authentication and encryption techniques relying on cryptographic primitives play crucial roles in guaranteeing reliable data, proper validation, and secure communication.[1-2, 8] However, most of nowadays cryptographic primitives are designed based on mathematical one-way functions such as Hash algorithm, prime number factorization, and discrete logarithm problem,[9-10] which have been demonstrated to be at risk from security breaches.[11] Thus, it is imperative to develop reliable methods capable of generating highly secure cryptographic primitives for information security.

Physical unclonable function (PUF), also known as physical one-way function, has been considered as one of the most promising hardware primitives in information security.[12-18].

Upon being interacted with external input challenges, PUFs will output corresponding responses, based on which intrinsic physical randomness could be extracted as fingerprints to distinguish each other.[12] Compared with conventional mathematical cryptographic primitives, PUF is fundamentally more secure due to the fact that it has nothing to do with any number theories and duplicating a PUF is almost impossible.[15] Benefiting from mature silicon technologies, various types of silicon-based semiconductor PUFs were widely investigated, such as arbiter PUF,[19] butterfly PUF,[20] flip-flop PUF[21] and ring oscillator PUF,[22] almost all of whose randomness are induced by inevitable uncontrollable random deviations during their manufacturing processes. However, these silicon PUFs only have a limited number of challenge-response pairs (CRPs) with low entropy,[14] usually leading to the demand of additional pre-processing or post-processing entropy compensation units which will bring about higher power consumption, area inefficiency, and expensive manufacturing costs. Other factors, including environmental variations, side-channel attacks, and hardware Trojans, also severely hinder the security of silicon-PUF based systems.[23-27] Therefore, scientists set their sights on new approaches such as nanotechnology, optics, and chemical methods to fabricate PUFs, while this context proposes several emerging PUFs based on nanowires,[28-29] nanofracture,[30] printed nanopatterns,[31] quantum dots,[32-33] surface wrinkles[34-36] carbon nanotubes,[14] plasmonic nanoparticles,[37-38] nano-electromechanical system,[39-40] and T cells.[26] Among these novel PUFs, optical PUF benefiting from its inherent huge complex optical random features,[15-16, 41] enjoys many superior properties such as high entropy, resource efficiency, and tamper proof, thus holding a promising potential for next-generation secure cryptographic primitives.[12, 42] For instance, Choi *et al.* proposed an optical cryptographic surface based on the photonic structure of micropillars array,[43] and Chen *et al.* reported a fluorescent PUF based on perovskite quantum dots/chaotic metasurfaces hybrid nanostructures.[33] However, the aforementioned functional structures of micropillars array and metasurfaces were obtained through traditional lithography or ion beam etching process, which is complicated, costly, and time-consuming. Therefore, more efficient, economical and novel alternative implementing approaches are imperative for desirable optical PUFs.

Nature has long time offered valuable materials and functional structures for human beings.[44-46] There are no two identical "leaves" in the world. The unique properties of biological structures open a new potential in the manufacture of PUF keys. Herein, we demonstrate a concept of bionic optical PUFs inspired from unique structures of natural plants. Our bionic optical PUFs are embodied in optical scattering films with random bionic micro-nano architectures molded from common plant tissues. Upon being illuminated by an input modulated laser light, the bionic PUF would generate a corresponding speckle response. Fundamental performances of the speckle responses are characterized by calculating inter-chip and intra-chip Hamming distances between binary codes converted from the speckle patterns, which reveal the excellent properties of bionic PUFs for being high-entropy, unique, robust, and of large encoding capacity, and thus bionic PUFs are favorable to be cryptographic primitives. Moreover, the applications of bionic PUFs in encrypted information communication and entity authentication are demonstrated. Since the manufacturing process of bionic PUFs is simple, low-cost, green and environmentally friendly, we believe the bionic PUFs could provide a new opportunity for safeguarding future information security.

## 2. RESULTS AND DISCUSSION

**2.1 Concept design, fabrication and morphology characterization.**

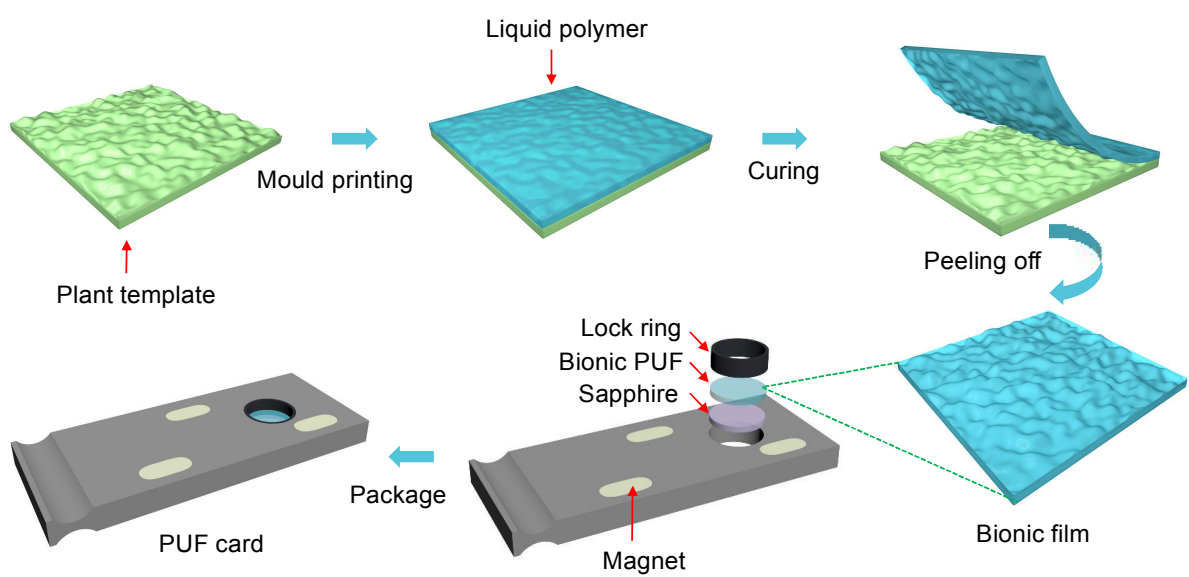

**Fig. 1.** Schematic of the process flow for manufacturing bionic film and PUF card.

Natural plants can be utilized as templates to produce the bionic structural films via a low-cost mould printing method, which is a type of soft lithography[47] and has been widely used in the field of flexible electronics, such as our previous researches.[48-49] Figure 1 depicts the fabrication process of bionic PUFs. First, the uncured polymer solution is uniformly coated onto the plant mold. After heat curing and peeling off, the bionic film with the negative surface structure of the plant is obtained. For avoiding mechanical abrasion and achieving functionalized application, the bionic film with a sapphire substrate and a lock ring is packaged into a card housing, thus obtaining a bionic PUF key in the shape of the rectangular card as displayed in the actual photos of Fig. S1 (ESI†).

In this study, two polymer materials including polydimethylsiloxane (PDMS) and polyvinyl alcohol (PVA) are applied to fabricate bionic PUF film, respectively. PDMS is a class of silicone with intriguing properties including the simple process, mechanical robustness, thermal stability, high transparency, biological compatibility, and chemical inertness.[6] Besides, PVA is a biodegradable and water-soluble polymer material and its aqueous solution can evaporate and solidify into a thin film, thus providing a candidate for processing micro-nano structured film via the soft lithography process. Here, we have studied four natural plant tissues with different surface structures and fabricated the related bionic films and PUF cards. The first plant tissue is the lotus leaf. Figure 2a shows a green lotus leaf with rolling drops of water. Behind the magical superhydrophobic property, the lotus leaf hides more wonderful morphology on its surface. There are several randomly distributed conical microtowers with a height of about 14 μm observed in the scanning electron microscopy (SEM) image as displayed in Figure 2b. More detailed fluffy nanostructures are also found on the conical microtowers. After the soft lithography process, a bionic polydimethylsiloxane (PDMS) film molded from lotus leaf is obtained. In general, flat pure PDMS film is transparent.[6] But the as-fabricated bionic PDMS film (Figure 2c) look like a haze, basically due to the unique architecture on its surface. Figure 2d reveals the false-color SEM image of lotus leaf-based bionic PDMS film with the spare and random circular cave. The diameters of caves are ranged from 4 μm to 13 μm and averaged on 7.8 μm (see Figure S1, Supporting Information). The interval distances of adjacent caves normally distribute from 9 μm to 30 μm, and the average value is 17.4 μm. Figure 2e shows the second plant tissue that is red rose petal symbolizing romance. The petal surface of the red rose possesses many mastoid-like structures with a height of ≈ 40 μm (Figure 2f), which is closely adjacent to each other. In this case, we fabricated the polyvinyl alcohol (PVA) film templated from rose petal, also with haze status in Figure 2g. The negative-mastoid structured caves are successfully printed, including the sharp edge of caves and the more specific wrinkles in the bottom of the caves (Figure 2h). Further, the top-view SEM image shows that the negative-mastoid caves of the bionic PVA film are tightly distributed with an average diameter of 25 μm, and interval gaps between the center points of neighboring caves coverage from 17 μm to 42 μm (see Figure S2, Supporting Information). In addition to rose petal, rose leaves also have featured dense island-like microstructures. The dependent bionic PDMS film shows an opaque state and contains microgrooves with an average inter-groove distance of ≈34 μm on the surface as placed in Figure S3 of Supporting Information. The fourth plant tissue is the ginkgo leaf. Bathed in the sunshine, the golden clusters of ginkgo leaves in Figure 2i look extraordinarily beautiful. There are long convex structures (Figure 2j) like continuous mountains on the surface of the ginkgo leaf. Furthermore, the haze bionic PDMS film molded from ginkgo leaf is displayed in Figure 2k. The continuous gully-like structures are randomly arrayed with the average interval distance of 27 μm and the coverage from 10 μm to 65 μm, as shown in Figure 2*l* and Figure S4 of Supporting Information.

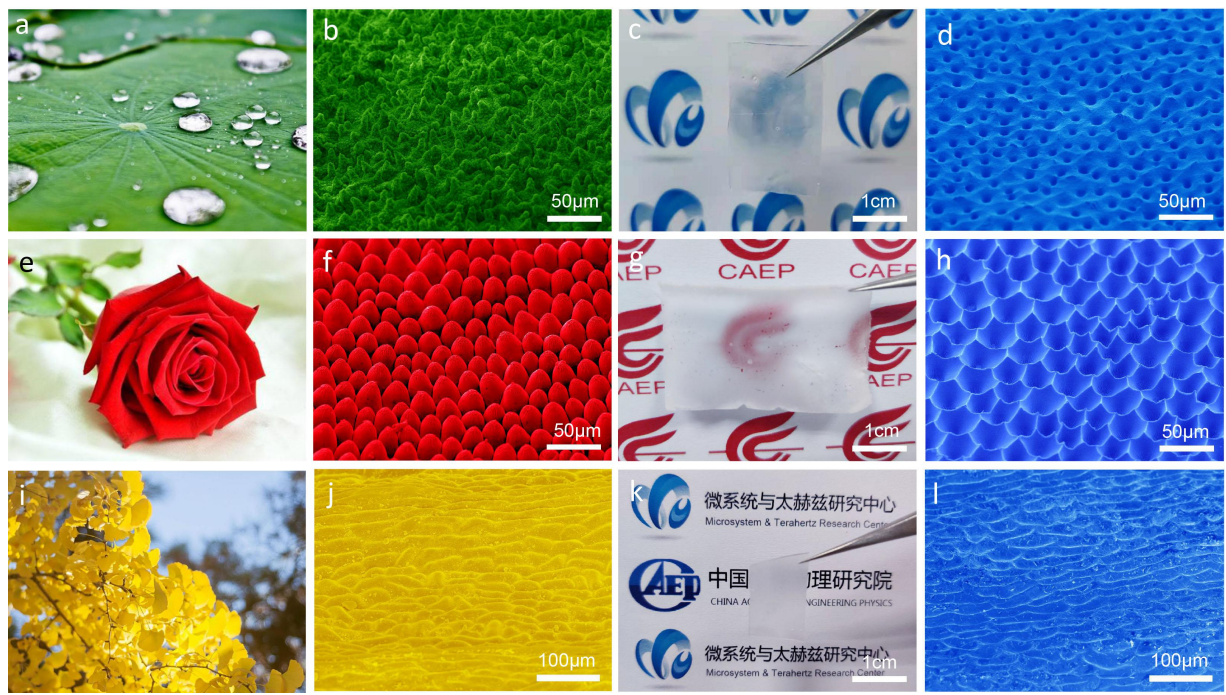

**Fig. 2**. Natural plants and bionic films. a) Photograph of a lotus leaf with rolling drops of water. b) A 45º titled view false-color SEM image of the lotus leaf. c) Photo image and d) surface morphology of the lotus leaf-based bionic PDMS film. e) Photograph of a red rose. f) False-color SEM image of the red rose petal. g) Photo image and h) 45º titled view false-color

SEM image of the rose petal based bionic PVA film. i) Photograph of a ginkgo biloba cluster. j) False-color SEM image of ginkgo leaf. k) Photo image and l) 45º titled view false-color SEM image of the as-fabricated ginkgo leaf-based bionic PDMS film.

**2.2 Basic performances of bionic PUFs**

**2.2.1. Observation of laser speckles.** Laser speckle has been commonly used in the research literature to fingerprint a wide range of objects for its convenience in 3D micro-nano physical character exploration.[12, 41-42] Here we will also analyze the performances of bionic PUFs through laser speckles. Figure 3a shows the experimental setup. A collimated and expanded Helium-Neon laser beam goes through the beam splitter and illuminates a liquid crystal spatial light modulator, on which random phase patterns are displayed (see Figure S5, Supporting Information) to create desirable challenges by modulating the laser beam wavefront through randomly setting each pixel's gray value. After penetrating through the polarizer, the lens and the iris, the input challenge is projected onto the front surface of the bionic PUF card, which is inserted in a card slot with precision < 1 μm. Behind the bionic PUF, the speckle patterns are detected by a CCD camera with 1280 × 1024 pixels. Figure 3b shows an observed speckle pattern. By filtered with the in-house developed algorithm of Gabor Hash,[12] the speckle pattern could be converted into two dimensional (2D) binary codes (Figure 3c) by assigning digital "1" to the "white" pixels and digital "0" to the "black" pixels. Equivalently, we can obtain a one dimensional (1D) key with a length ($L$) of 1280 Kbit.

**2.2.2. Evaluation of randomness, uniqueness and robustness**. In general, randomness, uniqueness and robustness are key parameters to evaluate the performance of PUFs for the requirements of cryptographic applications. Here we take lotus leaf as an example to evaluate the performances of bionic PUF.

For quantitative measure of the speckle randomness, entropies ($E$) along both $x$ and $y$ axes of the speckle pattern are obtained by using the standard definition of Shannon entropy as follow

$$E_i = -[p_i \log_2 p_i + (1-p_i) \log_2(1-p_i)] \qquad (1)$$

here $i = x$ or $y$, and $p_i$ is the probability of a bit being set to "1" along the respective axes.

Figure 3d shows the calculated results in which the average entropy is 0.995±0.002 and 0.996±0.003 along the *x* and *y* axis respectively, approaching the ideal value of 1 which indicates the perfectly random property of the PUF key's output.

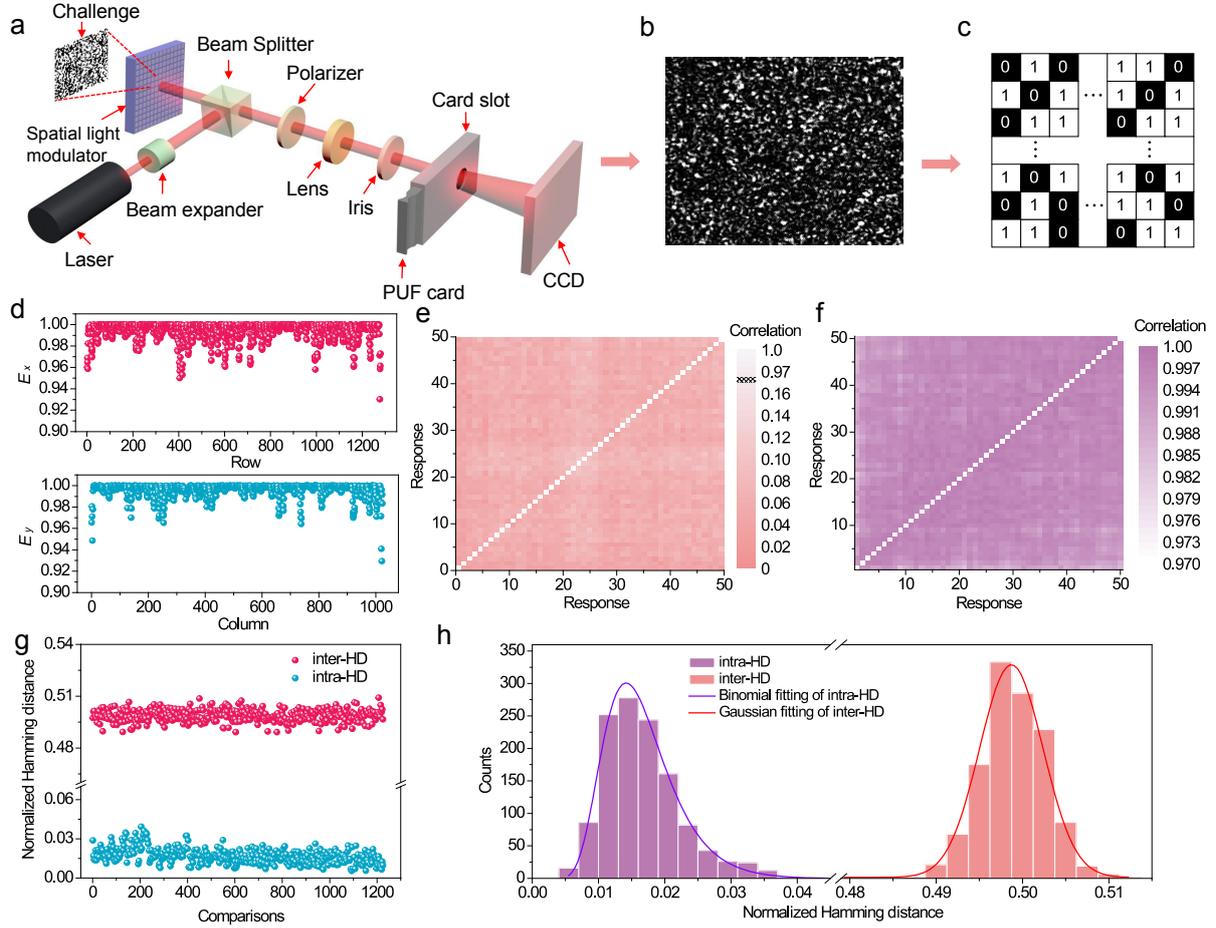

**Fig. 3.** Optical system for observation of laser speckles and performances of lotus leaf-based bionic PUF. a) Schematic of experimental setup. b) A speckle pattern generated by lotus leaf-based bionic PUF. c) Schematic of 2D binary codes extracted from speckle pattern. d) Entropies ($E_{x,y}$) along *x* and *y* axes respectively. e) Correlation coefficients among speckle patterns generated by 50 different PUFs with the same challenge. f) Correlation coefficients among 50 repeatedly tested speckle patterns. g) Inter-HDs between binarized speckle patterns generated by 50 different PUFs with the same challenge and intra-HDs obtained via repeatedly testing a single speckle pattern 50 times for one PUF card. h) Statistical distributions of the inter-HDs and the intra-HDs.

To assess the uniqueness and robustness of bionic PUFs, correlation coefficients (*C*) between response speckles could be statistically analyzed, which is described by the following

formula

$$C = \frac{\sum_m \sum_n (I_{mn}-\bar{I})(I'_{mn}-\bar{I'})}{\sqrt{(\sum_m \sum_n (I_{mn}-\bar{I})^2 \sum_m \sum_n (I'_{mn}-\bar{I'})^2}} \quad (2)$$

where $I_{mn}$ and $I'_{mn}$ are intensities at the $m_{th}$ row and the $n_{th}$ column of two speckle patterns, respectively, while $\bar{I}$ and $\bar{I'}$ are their corresponding average intensities. To evaluate the uniqueness of the speckle response, 50 different lotus leaf-based PUF cards were illuminated by the same challenge. Figure 3e shows the correlation coefficients among the detected speckle patterns. All the correlation coefficients are less than 10%, revealing that the speckle responses of different bionic PUFs are almost uncorrelated thus it is unique to each bionic PUF. To evaluate the robustness, we remeasure a single speckle pattern with PUF card repeatedly removing and reinserting into the card slot. The correlation coefficients among these repeatedly remeasured speckle patterns exceed 97% (Figure 3f), confirming qualified reproducibility and robustness of bionic PUF's response.

Hamming distance, which counts the bit-to-bit difference between two binary strings[12], can also be statistically analyzed to estimate the uniqueness and robustness of the bionic PUF. Typically, Inter-chip Hamming distance (inter-HD) can be used to assess uniqueness of PUFs, which describes differences between responses of different PUFs and is ideally 0.5.[18] Intra-chip Hamming distance (intra-HD), which assesses the robustness of the response signal of PUF, calculates the difference between repeatedly remeasured copies of a response of the same PUF with the same challenge and its value is ideally 0.[14] Figure 3g shows the statistical results of inter-HDs and intra-HDs, which were obtained via comparing binary codes converted from the speckle patterns used in Figure 3e and 3f respectively. The inter-HD shows a very narrow Gaussian distribution region from 0.485 to 0.51 with an average value of 0.499 and a variance of $5\times10^{-5}$, as shown in Figure 3h, demonstrating that responses of different PUFs are of significant difference and unique to their corresponding generating bionic PUFs. The degree of freedom ($F$) of the output binary codes can be calculated using the following equation[12, 50]

$$F = \mu(1-\mu)/\sigma^2 \quad (3)$$

where $\mu$ is the mean probability, and $\sigma^2$ is the variance. Remarkably, the binary output of the lotus leaf-based bionic PUFs shows a high degree of freedom up to 5000, resulting in a large binary coding space size of $2^{5000}$ ($\approx 1.4125 \times 10^{1505}$). Moreover, the intra-HDs show a very narrow Binomial distribution ranging from 0.005 to 0.038 with an average value of 0.016 (Figure 3h), approaching the ideal value of 0 and reinforcing the robustness claim for bionic PUF. The nonzero value of intra-HD mainly comes from misalignment of the PUF card, the vibrations of the optical system and the noises of the CCD camera. The fitting curves of the two distributions intersect at 0.388. This value can be regarded as a threshold to authenticate a candidate bionic PUF, which will be discussed in next subsection.

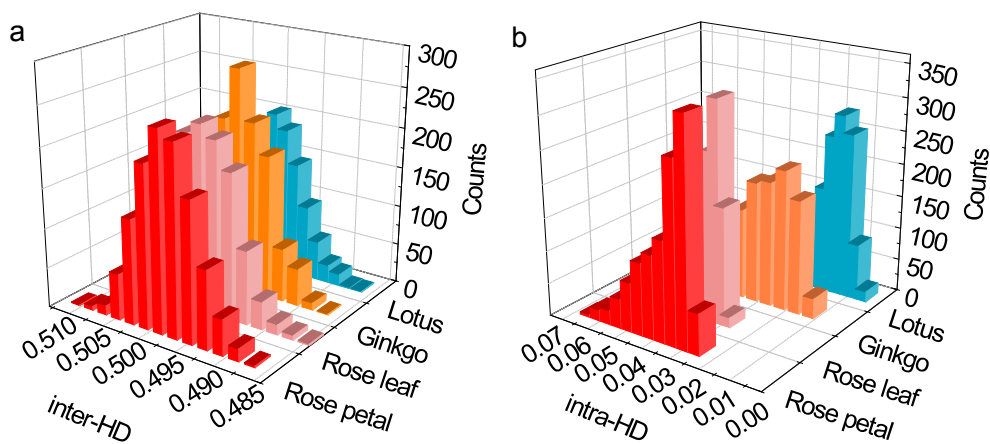

**Fig. 4.** Performances of bionic PUFs molded from different plant templates. a) Statistical distribution of the inter-HD for four kinds of bionic PUF molded from lotus, rose petal, rose leaf, and ginkgo, respectively. d) Statistical distribution of the intra-HD of these four kinds of bionic PUF outputs.

In addition to lotus leaf-based PUF, we also studied the other three kinds of bionic PUFs templated from rose petal, rose leaf and ginkgo leaf respectively. The inter-chip correlation coefficients of the bionic PUFs shown in Figure S6 (Supporting Information) indicate that different bionic PUFs are completely uncorrelated and unique. Figure 4a displays the statistical distribution of inter-HDs for the four types of bionic PUFs in this work. All the inter-HDs present a very narrow Gaussian distribution and are close to the ideal value of 0.5, once again confirming the uniqueness. Besides, repeated measurements of a single speckle response were also carried out for other bionic PUFs. As shown in Figure S7 (Supporting Information), the intra-chip response is highly similar, and the correlation coefficients are close to 100%. The intra-HD statistics for four kinds of bionic PUFs are placed in Figure 4b, revealing that all intra-HDs follow Binomial distribution with almost zero values less than

0.07 and demonstrating the desirable robustness of bionic PUF. Figure S8 (Supporting Information) shows the degree of freedom of the output binary code space of these four kinds of bionic PUFs, indicating that all of them have a large encoding capacity of at least $2^{3700}$. In view of the aforementioned high performances, we propose the notion of bionic PUFs and believe that it is applicable to many other natural materials that are not mentioned in this work.

## 2.3 Applications of bionic PUFs

**2.3.1 Entity authentication.** Identity authentication is critical in our daily life and industry production.[51-52] Given the advantages of randomness, uniqueness and robustness, we present a simple protocol based on the bionic optical PUFs for authentication applications, as shown in Figure 5a. The execution of authentication can be divided into two steps, including the registration and validation processes.

(i) In the registration stage, input challenges ($C_i$) are projected onto the surface of bionic PUFs to generate the associated speckle responses. The detected speckle patterns are transformed into binary keys ($K_i$). Then, the CRPs composed of the registered information of $C_i$ as well as $K_i$ are stored in the data cloud.

(ii) In the validation stage, the authenticator randomly selects $C_i$ from the data cloud to challenge the candidate PUF, and then obtain a check code ($K_i'$). Hereafter, the check code is uploaded to the data cloud to compare the Hamming Distance between $K_i$ and $K_i'$ ($HD(K_i, K_i')$). If $HD(K_i, K_i')$ is less than a preset threshold ($T$), the verification is true. Otherwise, the verification result is false.

In general, efficient authentication is an essential point for the whole validation process without any security and quality problem. To evaluate authentication reliability, the false accept rate (FAR) and false reject rate (FRR) can be analyzed by calculating the intersection areas of the two distribution fitting curves of inter-HDs and intra-HDs:

$$FAR = 1 - F(LT, L, d_1) \qquad (4)$$

$$FRR = F(LT, L, d_2) \qquad (5)$$

$$F(LT, L, d_i) = \sum_{j=0}^{LT} \binom{L}{j} d_i^j (1 - d_i)^{L-j} \qquad (6)$$

here, $L$ is the bit length of the key, $T$ is the abscissa of intersecting point between the two fitting curves ($T \approx 0.388$, as depicted in previous subsection), and $d_1$, $d_2$ are the mean values of intra-HD and inter-HD, respectively. By numeral calculation, we can arrive at a FAR and a FRR of both smaller than $10^{-200}$, which thus claims the ultra-high verification reliability of bionic PUF.

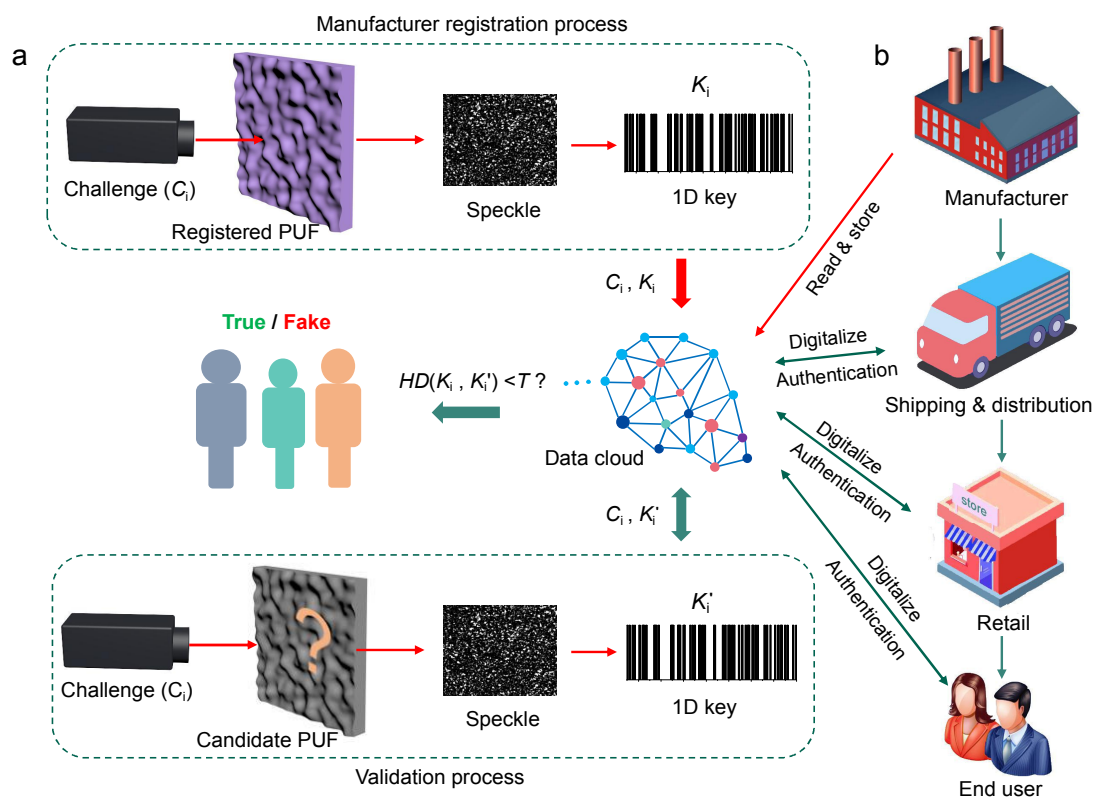

**Fig. 5.** An optical authentication system based on bionic PUFs. a) Schematic of registration and authentication processes. b) Schematic of the practical application of bionic PUF in the product verification covering the entire product supply chain.

To estimate the performance of the authentication protocol against thermal vibrations, we investigated the laser speckle responses of the PUF card under different temperature conditions ranging from the room temperature (26 °C) to 70 °C. Eventually, the obtained intra-HDs among the responses at different temperatures are always below the authentication threshold of 0.388 as shown in Fig. S10 (ESI†), indicating the desired reliability of the PUF card within a certain temperature range. It is worth noting that the speckles are sensitive to deformations when the flexible film suffers from the impacts of external forces, which thus restrict the functionalized application of the non-protected PUF films. Therefore, the card encapsulation reinforcement and slot locator (see Fig. S1, ESI†) have been exploited to avoid the physical abrasion of bionic films, and obtain ideal intra-HDs and robustness for long-time operation. From another point of view, the flexible property of bionic films also endows them with the potentiality as flexible PUF labels that can be compatibly attached and fixed onto some curved surfaces of entities in contrast to those of rigid PUFs. Such flexible PUF labels

**2.3.2 Encrypted communication.** Secure cryptographic primitives are highly desired for safeguarding information communication in the modern digital society.[15, 41] Traditional message encryption and communication between two parties need an additional resource of database to store and share the same secret keys, which is prone to malicious attacks. Herein, we design a protocol based on bionic optical PUF for the application in secure information communication. Message sender has a bionic PUF A as encrypted key $K_i(A)$. The listener holds another bionic PUF B as decrypted key $K_i(B)$. The two PUF keys are private and unknown to each other, and their XOR "key-mixture" is regarded as a public dictionary depicted in Figure 6a. The whole communication process (Figure 6b) can be divided into the encryption and decryption steps. In the sending stage, the message ($m$) is encrypted into a ciphertext of $m \oplus K_i(A)$, which is then transmitted to the listener via a public message transmission channel.

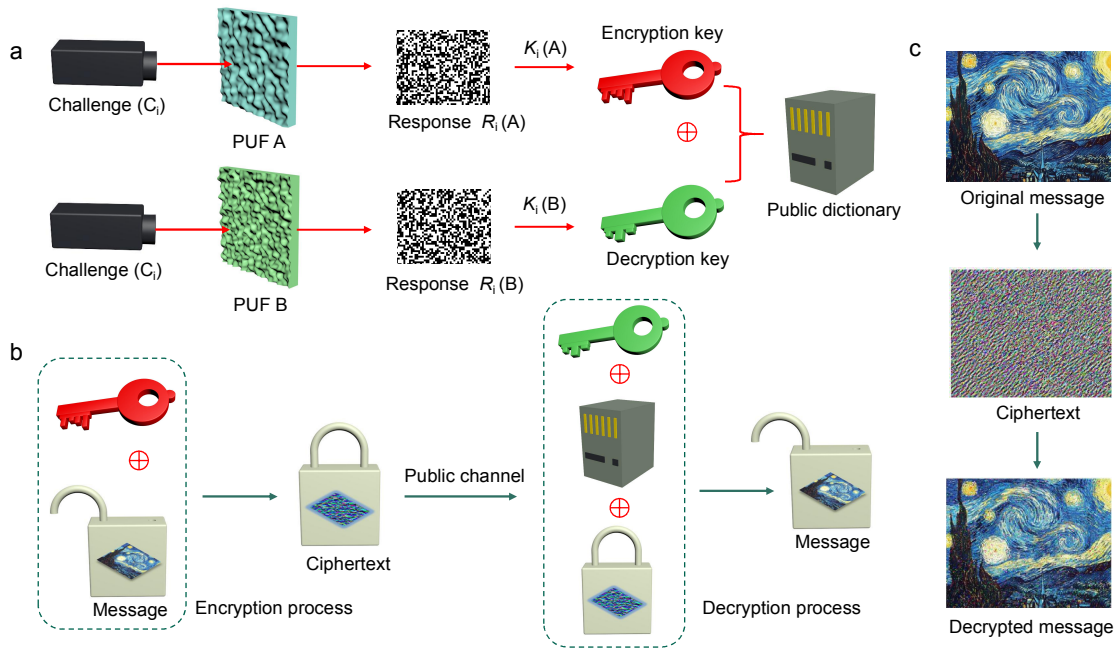

**Fig. 6.** Implementation protocol of bionic PUF based information encryption and transmission. a) Schematic of the generation of the encryption key and the decryption key. b) Mechanism of the encryption process and decryption process. c) A demo for encryption and transmission of painting *The Starry Night*.

As for the decryption stage, the listener utilizes the PUF B to decrypt the ciphertext with the following formula

$$K_i(B) \oplus [K_i(A) \oplus K_i(B)] \oplus (K_i(A) \oplus m) = m \qquad (7)$$

In the whole process of information communication, no extra stored resource is required, ensuring an efficient and secure transmission. Figure 6c displays a demo of message encryption and communication based on the protocol. It can be found that the painting *The Starry Night* is well encrypted, transmitted, and decrypted, which provides a promising potential in areas of national defense, information confrontation, financial transaction, and so forth.

## 3. CONCLUSIONS

In summary, we propose the concept of bionic PUFs inspired from biological diversity and uniqueness. Four types of optical bionic PUFs are fabricated by molding the surface structures of natural plant tissue such as lotus leaf, rose petal, rose leaf and ginkgo leaf. The manufacturing process is simple, low-cost, green and environmentally friendly. The statistical analysis of Hamming distances and correlation coefficients indicates that the extracted binary codes from speckle responses are random, unique and robust enough for all bionic PUFs to be widely applicable to cryptographic primitives. We also depict the entity authentication process based on the bionic PUF and illustrate the possible application in goods verification covering the entire product supply chain. Moreover, a protocol for information encryption and communication based on the bionic PUFs is proposed, which shows a functional demo in the practical application of information encryption and decryption. We expect that the notion of bionic PUFs may open a new pathway in future hardware-based information security.

## 4. EXPERIMENTAL DETAILS

**4.1. Fabrication of bionic PUFs.** In this study, we fabricated four kinds of bionic PUFs. Apart from that rose petal based bionic film is prepared by using PVA, the others based on the lotus leaf, rose leaf, and ginkgo leaf are all fabricated by utilizing PDMS. First, the plant tissue was cut into rectangular shapes and washed deionized water three times. After dried by $N_2$ gas blowing, the plant tissue template was fixed on a glass substrate using Scotch tapes. As for bionic PVA film, the PVA aqueous solution (0.1 g ml$^{-1}$) was uniformly coated on the surface of the rose petal template and placed at room temperature 25 °C for 24 hours. Then, cured bionic PVA film was pelt off for the next package of PUF cards. In regard to bionic PDMS film, the PDMS mixture liquid with a base to curing agent ratio of 10:1 was evenly poured onto the surface of the plant template. After curing at 70 °C for 1 hour, the PDMS film with the inverse structures of biological tissue was peeled off as bionic PUF film. Next, the bionic films were cut into round shapes (5 mm diameter) and packaged with transparent sapphire substrate and lock ring into rectangular card housing, thus obtaining bionic PUF key cards.

**4.2 Characterization setup.** The surface morphology of plant tissues and bionic films was investigated by optical microscope (Leica DM4 B), and scanning electron microscope (SEM, Carl Zeiss SUPRA 60) operated at 2.5 kV. The basic properties of bionic PUF were surveyed in our optical system stage depicted as the following step. (i) A collimated and expanded Helium-Neon laser beam goes through the beam splitter and illuminates a liquid crystal spatial light modulator, on which random phase patterns are displayed to create desirable challenges by modulating the laser beam wavefront through randomly setting each pixel's gray value. (ii) After penetrating through the polarizer, the lens and the iris, the input challenge projects onto the front surface of the bionic PUF card, which is inserted in a card slot. (iii) Behind the PUF card, the transmitted speckle pattern is detected by a CCD camera. (iv) Finally, the speckle patterns were filtered into binary codes by Gabor Hash, and the basic performances of bionic PUF were analyzed by MATLAB (The MathWorks, Inc) algorithm.

■ **ASSOCIATED CONTENT**

The Supporting Information is available free of charge on the ACS Publications website.

Additional details of the characterizations of the plant tissues and bionic films, challenge-response data of bionic PUFs are described in the text (PDF).

■ **AUTHOR INFORMATION**

**Corresponding Authors**

**Yao Yao -** Microsystem & Terahertz Research Center, China Academy of Engineering Physics,

Chengdu 610200, China; Institute of Electronic Engineering, China Academy of Engineering Physics, Mianyang 621999, China. Orcid: https://orcid.org/0000-0002-4160-1369; Email: yaoyao_mtrc@caep.cn

**Zhiguang Qiu** - School of Electronics and Information Technology, State Key Lab of Opto-Electronic Materials & Technologies, Guangdong Province Key Lab of Display Materials and Technologies, Sun Yat-sen University, Guangzhou 510275, China. Orcid: https://orcid.org/0000-0002-6198-1100; Email: zhiguangqiu88@gmail.com

**Authors**

**Yongbiao Wan** - Microsystem & Terahertz Research Center, China Academy of Engineering Physics, Chengdu 610200, China; Institute of Electronic Engineering, China Academy of Engineering Physics, Mianyang 621999, China.
Orcid: https://orcid.org/0000-0001-6101-464X

**Pidong Wang** - Microsystem & Terahertz Research Center, China Academy of Engineering Physics, Chengdu 610200, China; Institute of Electronic Engineering, China Academy of Engineering Physics, Mianyang 621999, China.

**Feng Huang** - Microsystem & Terahertz Research Center, China Academy of Engineering Physics, Chengdu 610200, China; Institute of Electronic Engineering, China Academy of Engineering Physics, Mianyang 621999, China.

**Jun Yuan** - Microsystem & Terahertz Research Center, China Academy of Engineering Physics, Chengdu 610200, China; Institute of Electronic Engineering, China Academy of Engineering Physics, Mianyang 621999, China.

**Dong Li** - Microsystem & Terahertz Research Center, China Academy of Engineering Physics, Chengdu 610200, China; Institute of Electronic Engineering, China Academy of Engineering Physics, Mianyang 621999, China.

**Kun Chen** - Microsystem & Terahertz Research Center, China Academy of Engineering Physics, Chengdu 610200, China; Institute of Electronic Engineering, China Academy of Engineering Physics, Mianyang 621999, China.

**Jianbin Kang** - Microsystem & Terahertz Research Center, China Academy of Engineering Physics, Chengdu 610200, China; Institute of Electronic Engineering, China Academy of Engineering Physics, Mianyang 621999, China.

**Qian Li** - Microsystem & Terahertz Research Center, China Academy of Engineering Physics, Chengdu 610200, China; Institute of Electronic Engineering, China Academy of Engineering Physics, Mianyang 621999, China.

**Taiping Zhang** - Microsystem & Terahertz Research Center, China Academy of Engineering Physics, Chengdu 610200, China; Institute of Electronic Engineering, China Academy of Engineering Physics, Mianyang 621999, China.

**Song Sun** - Microsystem & Terahertz Research Center, China Academy of Engineering Physics, Chengdu 610200, China; Institute of Electronic Engineering, China Academy of Engineering Physics, Mianyang 621999, China.

**Author Contributions**

[#] Y. Wan and P. Wang contributed equally to this work. Y. Wan and Y. Yao conceived the idea. Y. Yao guided the project. Y. Wan, P. Wang and Z. Qiu implemented main experiments. F. Huang did the package of PUF cards. J. Yuan, D. Li, K. Chen, J. Kang, Q. Li, T. Zhang and S. Sun participated in the data analysis and figures layout. Y. Wan, P. Wang, Z. Qiu and Y. Yao draft the manuscript. All authors contributed to the revised manuscript.

**Notes**

The authors declare no competing financial interest

■ **ACKNOWLEDGMENTS**

This study was financially supported by the funds of the Science Challenging Project (No. TZ2018003) and the National Natural Science Foundation of China (61875178, 61805218). Y. Wan would also like to thank Dr. Ping Jin from West China Medical Center of Sichuan University for her supply of gingko leaves. Could you spend the rest of your life with me?

■ **REFERENCES**

# Supporting Information

# Bionic Physical Unclonable Functions

Yongbiao Wan[†,‡,#], Pidong Wang[†,‡,#], Feng Huang[†,‡], Jun Yuan[†,‡], Dong Li[†,‡], Kun Chen[†,‡], Jianbin Kang[†,‡], Qian Li[†,‡], Taiping Zhang[†,‡], Song Sun[†,‡], Zhiguang Qiu[§,*], Yao Yao[†,‡,*]

[†] Microsystem & Terahertz Research Center, China Academy of Engineering Physics, Chengdu 610200, China
[‡] Institute of Electronic Engineering, China Academy of Engineering Physics, Mianyang 621999, China
[§] School of Electronics and Information Technology, State Key Lab of Opto-Electronic Materials & Technologies, Guangdong Province Key Lab of Display Materials and Technologies, Sun Yat-sen University, Guangzhou 510275, China.

* To whom correspondence should be addressed. Email: zhiguangqiu88@gmail.com; yaoyao_mtrc@caep.cn

[#] These authors contributed equally to this work.

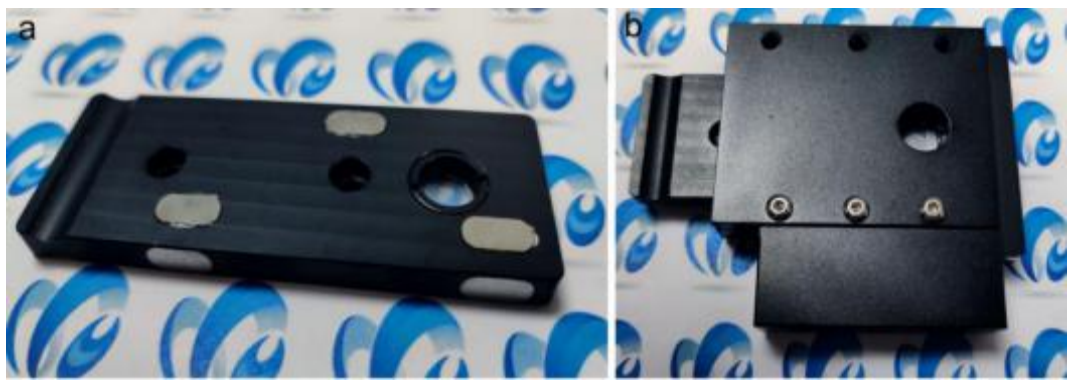

**Fig. S1.** The actual photos of a) a bionic PUF card and b) the slot inserted card.

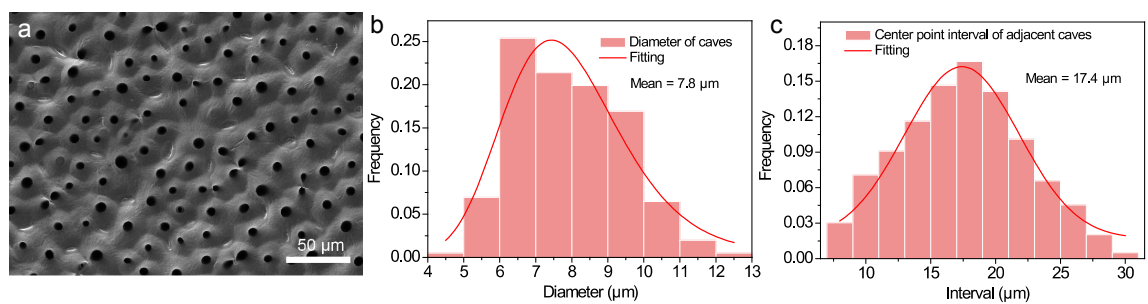

**Fig. S2.** a) Top view SEM image of lotus leaf based PDMS film. b) Statistical distribution of diameter of the caves in panel a, showing the average diameter of 7.8 μm and the distribution range from 4 μm to 13 μm. c) Statistical distribution of interval distance of the adjacent caves in panel a, demonstrating the average interval gap of 17.4 μm and the coverage from 9 μm to 30 μm.

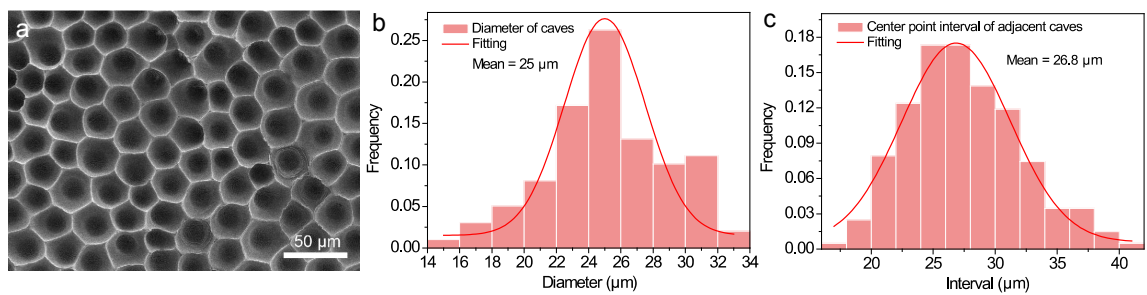

**Fig. S3.** a) Top view SEM image of rose petal based PVA film. b) Statistical distribution of diameter of the caves in panel a, showing the average diameter of 25 μm and the distribution range from 14 μm to 34 μm. c) Statistical distribution of center point interval distance of the adjacent caves in panel a, demonstrating the average interval gap of 26.8 μm and the coverage from 17 μm to 42 μm.

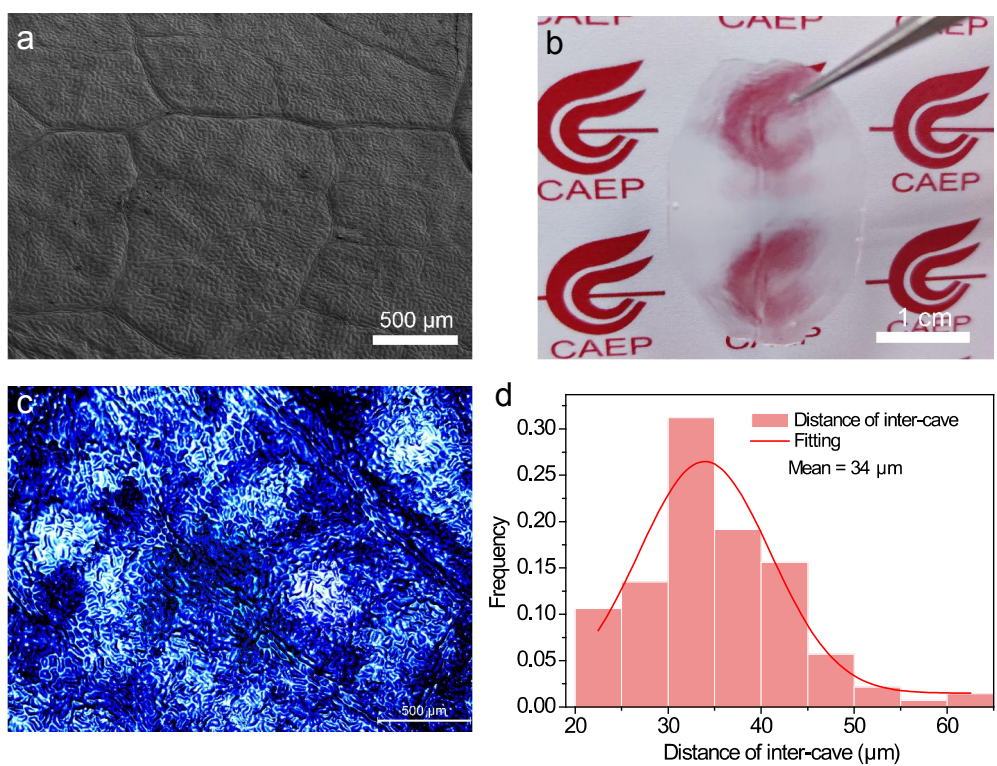

**Fig. S4.** a) SEM image of rose leaf. b) Photograph and c) top view optical morphology of rose leaf based bionic PDMS film. d) Statistical distribution of the distance of inter-cave in panel c, indicating the average interval gap of 34μm and the coverage from 20 μm to 65 μm.

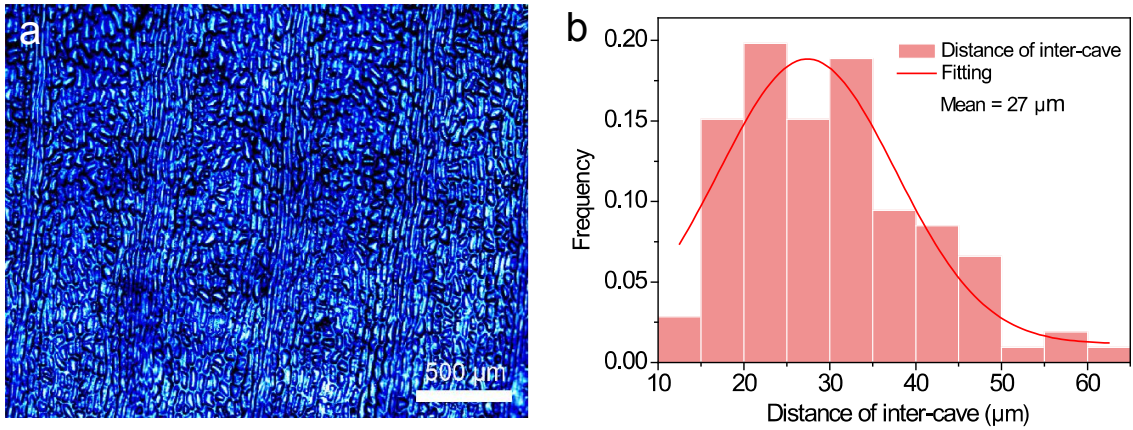

**Fig. S5.** a) Top-view optical morphology of ginkgo leaf based-bionic PDMS film. b) Statistical distribution of the distance of inter-cave in panel a, showing the average interval distance of 27 μm and the coverage from 10 μm to 65 μm.

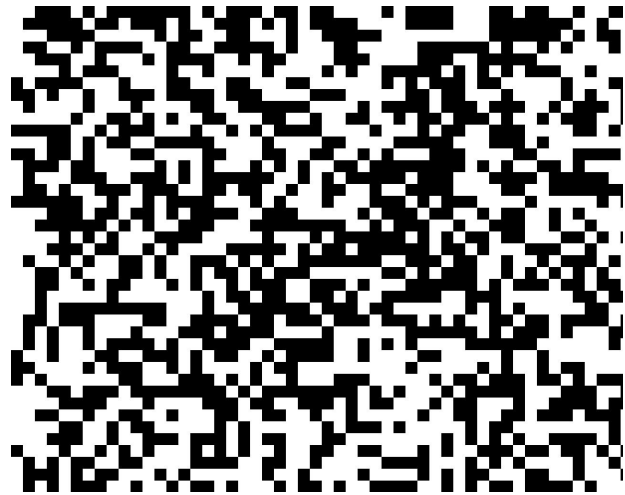

**Fig. S6.** A random phase pattern for modulating the input laser beam, which creates a desirable challenge.

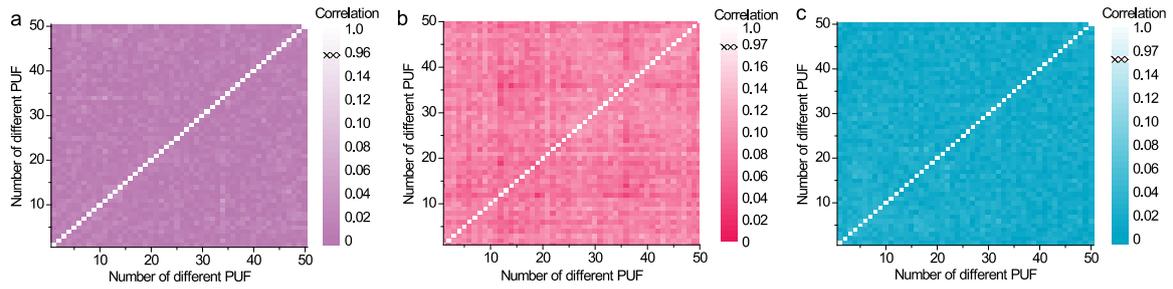

**Fig. S7.** Correlation coefficients of inter-chip PUFs based on a) rose petal, b) rose leaf, and c) ginkgo leaf, respectively. Inter-chip means that 50 different bionic PUFs were tested and compared with each other.

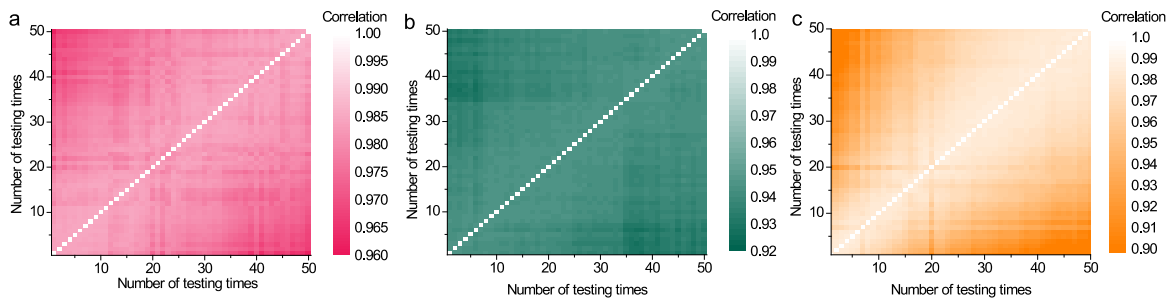

**Fig. S8.** Correlation coefficients of intra-chip PUFs based on a) rose petal, b) rose leaf, and c) ginkgo leaf, respectively. Intra-chip presents that one PUF card was repeatedly tested over 50 times.

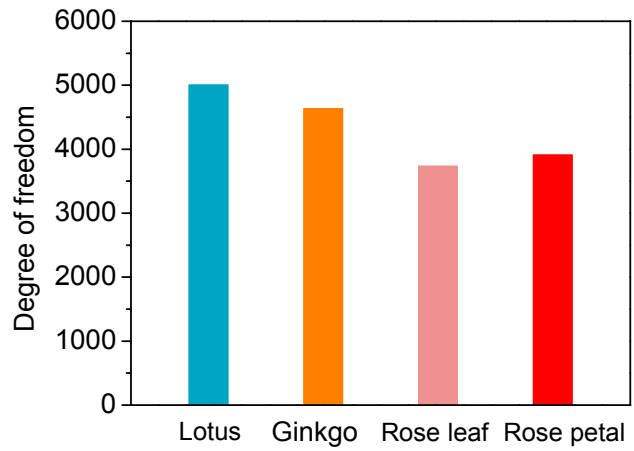

**Fig. S9.** Degree of freedom of these four kinds of bionic PUFs outputs.

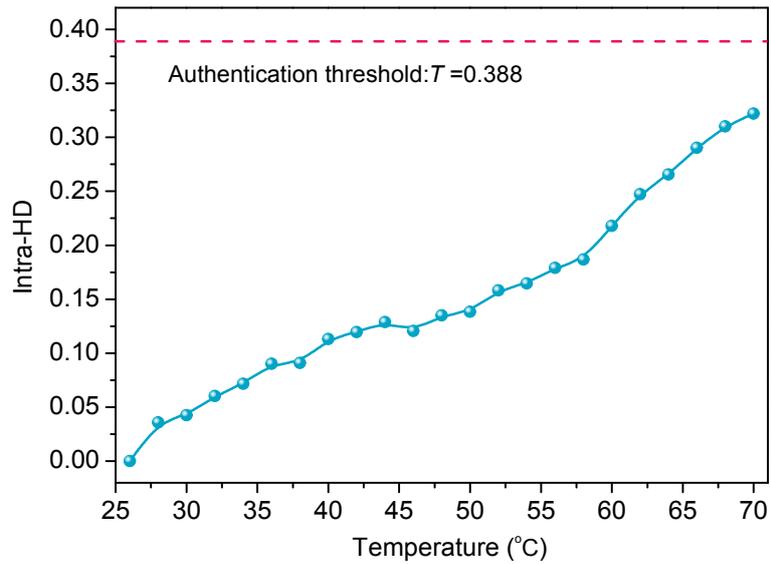

**Fig. S10.** Performance of the authentication protocol against thermal vibration. Intra-HDs ascend with increasing temperature while maintain below authentication threshold of 0.388, showing the reliability of PUF card within a certain temperature range.